\documentclass[reprint,superscriptaddress,amsmath,amssymb,aps,pra]{revtex4-2}
\usepackage[colorlinks, linkcolor = ForestGreen, anchorcolor = blue, citecolor = magenta]{hyperref}
\usepackage{amsmath,amssymb,amsthm,mathtools,mathrsfs}
\usepackage[table,dvipsnames]{xcolor}
\usepackage{tikz}
\usepackage{adjustbox}
\usepackage{dsfont}
\usepackage[normalem]{ulem}
\usepackage{enumerate}
\usepackage[all]{xy} 
\usepackage{physics}

\theoremstyle{plain}
\newtheorem{theorem}{Theorem}
\newtheorem{lemma}{Lemma}
\newtheorem{proposition}{Proposition}

\theoremstyle{definition}
\newtheorem{definition}{Definition}
\newtheorem{remark}{Remark}
\newtheorem{example}{Example}

\newcommand{\bd}{\begin{definition}}
\newcommand{\ed}{\end{definition}}
\newcommand{\bt}{\begin{theorem}}
\newcommand{\et}{\end{theorem}}
\newcommand{\bn}{\begin{proposition}}
\newcommand{\en}{\end{proposition}}
\newcommand{\be}{\begin{equation}}
\newcommand{\ee}{\end{equation}}
\newcommand{\blem}{\begin{lemma}}
\newcommand{\elem}{\end{lemma}}
\newcommand{\bx}{\begin{example}}
\newcommand{\ex}{\end{example}}
\newcommand{\bprf}{\begin{proof}}
\newcommand{\eprf}{\end{proof}}

\DeclareMathAlphabet{\mathpzc}{OT1}{pzc}{m}{it} 
 \DeclareFontFamily{OT1}{pzc}{}
 \DeclareFontShape{OT1}{pzc}{m}{it}{ <-> s*[1.2] pzcmi7t }{}
 \DeclareMathAlphabet{\mathpzc}{OT1}{pzc}{m}{it}
 
\def\E{\mathcal{E}}
\def\H{\mathcal{H}}

\begin{document}									
\preprint{APS/123-QED}

\title{Passive realism in the presence of open system dynamics}

\author{James Fullwood}
\email{fullwood@hainanu.edu.cn}
\affiliation{School of Mathematics and Statistics, Hainan University, Haikou, Hainan Province, 570228, China}
\affiliation{Hainan International Exchange Center for Theoretical Physics, Haikou, Hainan Province, 570228, China}
\author{Boyu Yang}
\email{byy@hainanu.edu.cn}
\affiliation{School of Mathematics and Statistics, Hainan University, Haikou, Hainan Province, 570228, China}
\author{Weixiang Ye}
\email{wxy@hainanu.edu.cn}
\affiliation{Hainan International Exchange Center for Theoretical Physics, Haikou, Hainan Province, 570228, China}
\affiliation{Center for Theoretical Physics, School of Physics and Optoelectronic Engineering, Hainan University, Haikou, Hainan Province, 570228, China}

\date{\today}

\begin{abstract}
We introduce the terminology \emph{passive realism} to encapsulate the scale-independent assumption that a physical system's pre-existing properties can be passively observed without altering its subsequent dynamics. A necessary condition for passive realism is provided by the no-signalling in time (NSIT) condition, which states that a measurement outcome is independent of whether or not a prior measurement was performed. While closed quantum systems generically violate NSIT due to measurement back action, we investigate whether open-system dynamics can dissipate this disturbance to the environment and restore NSIT. By modeling the time evolution between sequential measurements as an arbitrary quantum channel, we show that such environmental masking is possible only in two special cases. Specifically, we show that if the initial state of a two-time sequential measurement scenario is not maximally mixed, and if the channel governing the dynamics between measurements is not a discard-and-prepare channel, then there necessarily exist binary projective measurements violating NSIT, hence violating passive realism. By providing a constructive procedure for determining these violating measurements---applied here to depolarizing and dephasing channels---we demonstrate that, outside these two cases, system-environment interactions cannot erase the statistical footprint of a quantum measurement for all measurement choices.
\end{abstract}

	\maketitle
	
\section{Introduction}

In the 1980s, Leggett and Garg introduced the concept of ``macroscopic realism" to probe the boundary between the quantum and classical worlds~\cite{LeGa85}. In particular, they were testing whether genuinely macroscopic systems could exist in coherent quantum superpositions. Over the intervening decades, however, this terminology has experienced significant semantic drift. Today, Leggett-Garg inequalities and related tests of ``macrorealism" are routinely applied to manifestly microscopic entities, such as single photons or isolated nuclear spins. When applied to such systems, the assumption being tested is simply the conjunction of two postulates, namely, that a physical system possesses definite, pre-existing properties independent of measurement (realism), and that these properties can be passively observed without altering the system's subsequent dynamics (non-invasive measurability). As these postulates apply equally to microscopic and macroscopic systems alike, we introduce the terminology \emph{passive realism} to describe a scale-independent framework governed strictly by these two assumptions.

Leggett-Garg inequalities provide a well-known necessary condition for passive realism~\cite{Emary_2013,Halliwell_2017}. However, their formulation requires measurements at a minimum of three distinct times. In contrast, the no-signalling in time (NSIT) condition introduced by Kofler and Brukner offers a more streamlined approach, as it provides a necessary condition for passive realism using only a two-time sequential measurement scenario~\cite{Kofler_2013}. Physically, NSIT demands that the probability of a measurement outcome is completely independent of whether or not an earlier measurement was performed. While it is well-established that NSIT is generically violated for quantum systems which evolve unitarily between measurements due to measurement back action, here we consider violations of NSIT in the presence of system-environment interactions~\cite{Friedenberger_2017,Knee_2018}, where the evolution between measurements is modeled by a general quantum channel rather than a unitary operator. In such a context, a necessary and sufficient condition for violation of NSIT was established in Ref.~\cite{Comar_2026} for a fixed pair of measurements to be performed in sequence, and Ref.~\cite{Bartkiewicz_2026} establishes general results on NSIT violations for mutually unbiased measurements. Our results, in contrast, are inherently measurement-independent. In particular, we explicitly determine which initial states and open-system dynamics universally satisfy NSIT across \emph{all} possible projective measurements. 

By adopting the standard quantum formalism, one accepts the violation of non-invasive measurability as a foundational premise, dictated by the state-update rule following a measurement. Therefore, demonstrating a failure of passive realism for a closed quantum system is, in a sense, theoretically predetermined. As such, the fundamental question we address here is not whether measurement back action exists, but whether system-environment interactions can erase its statistical footprint at the system output. If such environmental masking were perfectly effective, an emergent passivity of measurements could be realized, thus restoring NSIT. However, we show that open-system dynamics can universally restore NSIT only in the two special cases. Specifically, we prove that NSIT is universally satisfied if and only if the initial state of a sequential measurement scenario is maximally mixed, or if the quantum channel governing the dynamics between measurements is a discard-and-prepare channel, whose output state is independent of its input.

Consequently, if the initial state of a two-time sequential measurement scenario is not maximally mixed, and if the dynamics between measurements are not modeled by a discard-and-prepare channel, then there necessarily exist measurements violating NSIT. This implies that passive realism is generically violated for sequential measurements, even in the presence of open-system dynamics. Moreover, we provide a direct constructive procedure to determine these violating measurements, which we show can always be taken to be simple dichotomic observables.

As an illustration of our results, we consider a single qubit subjected to both depolarizing and dephasing dynamics between measurements. In the depolarizing case, we construct a family of NSIT-violating measurements depending on the depolarization parameter. In accordance with our results, such measurements cease to exist once the channel becomes completely depolarizing. In contrast, as a dephasing channel preserves classical populations, it never becomes a discard-and-prepare channel. As such, we are able to construct a family of NSIT-violating measurements for all values of the dephasing parameter, explicitly confirming that even complete dephasing does not guarantee universal NSIT.

\section{No-signalling in time}

We consider a two-time sequential measurement scenario consisting of the following protocol: (i) Alice prepares a fixed state $\rho_A$ of a quantum system $A$ with finite-dimensional Hilbert space $\H_A$ of dimension $d_A>1$. (ii) Alice performs a projective measurement $\{P_i\}$ on system $A$ (so that $P_iP_j=\delta_{ij}P_i$ and $\sum_i P_i = \mathds{1}_A$). (iii) Following this initial measurement, the updated system evolves according to a completely positive trace-preserving linear map $\E:\mathcal{L}(\H_A)\to \mathcal{L}(\H_B)$, where $\mathcal{L}(\H)$ denotes the algebra of linear operators on a Hilbert space $\H$. Such a map $\E$ will be referred to as a \emph{quantum channel}. (iv) Bob performs a projective measurement $\{Q_j\}$ on the output of the channel $\E$ (which is a state on system $B$).

\begin{figure}[t]
    \centering
    \includegraphics[width=\columnwidth]
    {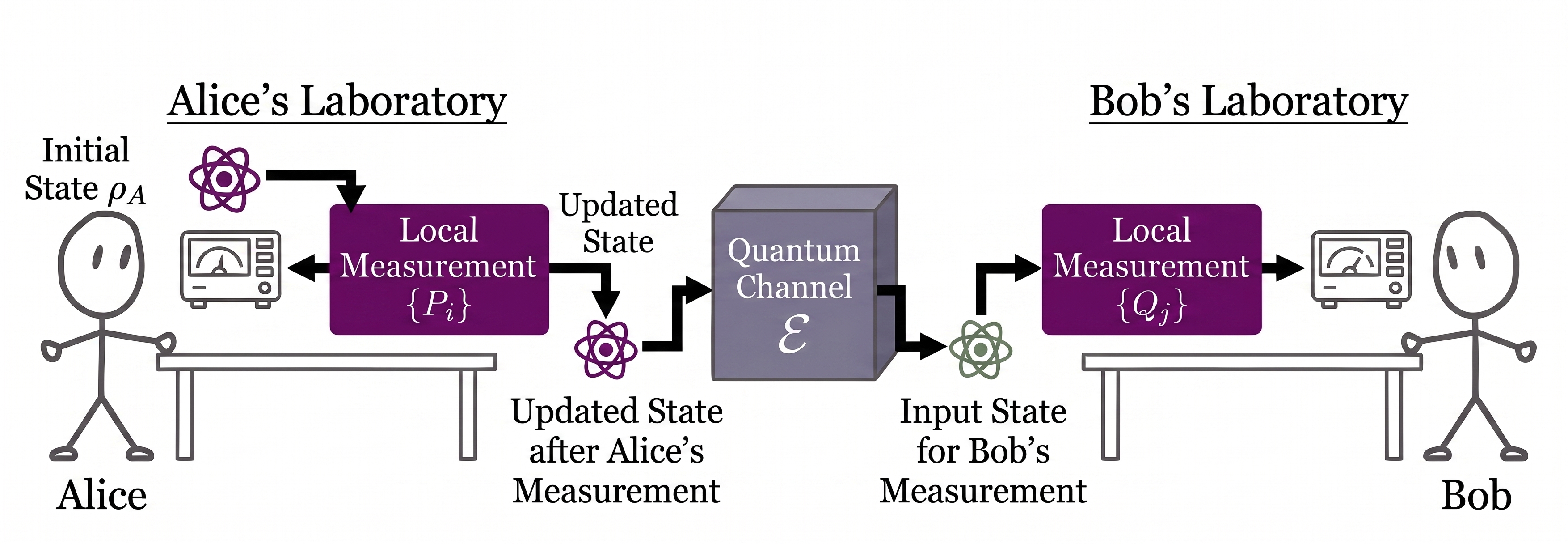}
    \caption{
  A two-time sequential measurement scenario of an open quantum system.}
    \label{fig_1}
\end{figure}

In such a two-time sequential measurement scenario (as depicted in Figure~\ref{fig_1}), it follows from the L\"{u}ders-von~Neumann projection postulate~\cite{Lu06} that the joint probability $\text{Prob}_{AB}(i, j)$ of Alice obtaining outcome $P_i$ followed by Bob obtaining outcome $Q_j$ is given by
\be \label{LvN17}
\text{Prob}_{AB}(i, j) = \Tr (\E(P_i \rho_A P_i)Q_j ) \, ,
\ee
which we refer to as the \emph{L\"{u}ders-von~Neumann (LvN) distribution}. In such a case, it follows that the probability $\text{Prob}_B(j)$ that Bob obtains outcome $Q_j$ is then given by
\be \label{BOBXPRX17}
\text{Prob}_B(j)= \sum_i \Tr (\E(P_i \rho_A P_i)Q_j)\, .
\ee
Bob's distribution $\text{Prob}_B$ is then said to satisfy the \emph{no-signalling in time} (NSIT) condition if and only if it is independent of whether or not Alice actually performs a measurement. As the trivial projective measurement $\{\mathds{1}_A\}$ corresponds to Alice not performing a measurement, it follows from Eq.~\eqref{BOBXPRX17} that NSIT translates to the mathematical condition that for all $j$ we have
\be \label{NSIT57}
\sum_i \Tr (\E(P_i \rho_A P_i)Q_j )=\Tr(\E(\rho_A)Q_j)\, .
\ee

\begin{remark}
It is important to note that we do not assume the projection operators constituting a projective measurement are of rank one. Moreover, we assume that a projective measurement $\{P_i\}$ is instantiated by the L\"{u}ders instrument $\{\mathcal{M}_i\}$ given by $\mathcal{M}_i(\rho_A)=P_i\rho_A P_i$. The L\"{u}ders instrument then guarantees that the state is updated without any extraneous unitary mixing within the subspace it is projected onto, thereby representing a minimally disturbing implementation of the measurement. This instrument-level specification is relevant to the relation between NSIT and quantum nondisturbance~\cite{Uola_2019}.
\end{remark}

\begin{remark}
By summing the L\"{u}ders-von~Neumann distribution \eqref{LvN17} over all of Bob's possible outcomes, the fact that $\E$ is trace-preserving implies that the probabilities $\text{Prob}_A(i)$ for Alice's individual measurement outcomes are given by $\text{Prob}_A(i)=\Tr(\rho_A P_i)$. As such, it follows that there are no retrocausal effects in such a sequential measurement scenario. In particular, Alice's measurement outcomes are completely independent of which measurement Bob performs. This is often referred to as the `arrow of time' (or `causality') condition in the literature~\cite{Leggett_2002,Chiribella_2011,Kofler_2013,Clemente_2015,Vitag_2023}.
\end{remark}

To see why passive realism necessitates the NSIT condition~\eqref{NSIT57}, we can isolate the distinct mathematical implications of its two defining postulates. By the assumption of realism, system $A$ possesses definite, pre-existing properties independent of whether or not such properties are measured. Therefore, Bob's probability of obtaining outcome $Q_j$ can be expressed as a marginal distribution over Alice's possible measurement outcomes regardless of whether or not Alice performs a measurement. As such, we have $\text{Prob}_B(j) = \sum_i \text{Prob}_{AB}(i,j \,|\, \text{Alice measures})$. On the other hand, the assumption of non-invasive measurability dictates that the physical act of Alice measuring these properties does not alter the system's subsequent dynamics. This implies that Bob's outcome probabilities remain exactly the same even if Alice's measurement is omitted, yielding $\text{Prob}_B(j) = \text{Prob}_B(j \,|\, \text{Alice does not measure})$. Passive realism then implies $\sum_i \text{Prob}_{AB}(i,j \,|\, \text{Alice measures}) = \text{Prob}_B(j \,|\, \text{Alice does not measure})$ for all $j$, which is precisely the NSIT condition~\eqref{NSIT57}.

\section{The main result}

There are several sufficient conditions for the NSIT condition \eqref{NSIT57} which are immediate. In particular, if $[\rho_A,P_i]=0$ for all $i$, if $[P_i,\E^{\dag}(Q_j)]=0$ for all $i$ and $j$ (where $\E^{\dag}$ is the Hilbert-Schmidt adjoint of $\E$), or if $\E$ is a discard-and-prepare channel---so that there exists a state $\sigma_B$ of $B$ such that $\E(\omega_A)=\sigma_B$ for all states $\omega_A$ of $A$--- then the NSIT condition \eqref{NSIT57} is satisfied. It then follows that if $\rho_A=\mathds{1}_A/d_A$ is maximally mixed, or if $\E$ is a discard-and-prepare channel, then the NSIT condition \eqref{NSIT57} is satisfied \emph{for every} possible projective measurement Alice and Bob can perform. In Ref.~\cite{FuWu_2026}, it was conjectured that the converse to this statement also holds, which we now answer in the affirmative:

\bt \label{MXT71}
If NSIT is satisfied for every projective measurement Alice and Bob can perform, then $\rho_A=\mathds{1}_A/d_A$ is maximally mixed, or $\E$ is a discard-and-prepare channel. 
\et

We note that the contrapositive of Theorem~\ref{MXT71} states that if $\rho_A$ is \emph{not} maximally mixed, and if $\E$ is \emph{not} a discard-and-prepare channel, then there necessarily exist projective measurements $\{P_i\}$ and $\{Q_j\}$ such that NSIT is violated, thus implying a failure of passive realism. Moreover, we will show later on that in such a case there always exist binary measurements $\{P,\mathds{1}_A-P\}$ and $\{Q,\mathds{1}_B-Q\}$ violating NSIT. 

Before proving Theorem~\ref{MXT71}, we establish a fundamental structural property regarding the state disturbances induced by projective measurements. For this, let $\text{Herm}_0(\H_A)$ denote the real vector space of traceless Hermitian operators on $\H_A$. For a given rank-one projector $P$, the nonselective L\"{u}ders disturbance on a state $\rho_A$ relative to performing no measurement is $\Delta_P(\rho_A)=\rho_A - P\rho_A P - (\mathds{1}_A - P)\rho_A(\mathds{1}_A - P)$, which can be expressed compactly as the double commutator $\Delta_P(\rho_A) = [P, [P, \rho_A]]$. 

\blem \label{lem:span}
Let $\rho_A$ be a state on $\H_A$. If $\rho_A$ is not the maximally mixed state $\mathds{1}_A/d_A$, the real linear span of the disturbances $\Delta_P(\rho_A)$ over all rank-one projectors $P$ is exactly the full space of traceless Hermitian operators, $\emph{Herm}_0(\H_A)$.
\elem

\bprf
Let $W_{\rho_A} = \text{span}_{\mathbb{R}}\{ \Delta_P(\rho_A) \mid \text{rank}(P)=1 \}$. Since $\Delta_P(\rho_A)=[P,[P,\rho_A]]$ is Hermitian and traceless for all rank-one projectors $P$, we have $W_{\rho_A} \subseteq \text{Herm}_0(\H_A)$. To show $W_{\rho_A} = \text{Herm}_0(\H_A)$, we suppose an operator $X \in \text{Herm}_0(\H_A)$ is orthogonal to $W_{\rho_A}$ with respect to the Hilbert-Schmidt inner product, and show that this necessitates $X=0$. 

Orthogonality implies $\Tr(X [P_v,[P_v,\rho_A]]) = 0$ for every rank-one projector $P_v = \dyad{v}{v}$. Utilizing the cyclicity of the trace, this condition reduces to
\be \label{eq:ye_ortho}
\langle v | \{X, \rho_A\} | v \rangle = 2 \langle v | X | v \rangle \langle v | \rho_A | v \rangle
\ee
for all unit vectors $|v\rangle \in \H_A$. Let $\rho_A = \sum_\lambda \lambda \Pi_\lambda$ be the spectral decomposition of $\rho_A$ over its distinct eigenvalues. Since $\rho_A \neq \mathds{1}_A/d_A$, there exist at least two distinct eigenvalues $\lambda$ and $\mu$ of $\rho_A$. Choose normalized vectors $|x\rangle \in \text{Im}(\Pi_\lambda)$ and $|y\rangle \in \text{Im}(\Pi_\mu)$. For real parameters $\theta$ and $\phi$, we define the unit vector $|v\rangle = \cos\theta |x\rangle + e^{i\phi} \sin\theta |y\rangle$. Substituting $|v\rangle$ into Eq.~\eqref{eq:ye_ortho} and isolating the terms dependent on $\theta$ and $\phi$ yields the requirement $(\lambda - \mu)(\langle x|X|x\rangle - \langle y|X|y\rangle) = 0$ and $(\lambda - \mu)\text{Re}(e^{i\phi}\langle x|X|y\rangle) = 0$. Since $\lambda \neq \mu$, this implies $\langle x|X|x\rangle = \langle y|X|y\rangle$ and $\langle x|X|y\rangle = 0$. 

Since $|x\rangle$ and $|y\rangle$ were arbitrarily chosen from distinct eigenspaces, $\langle v|X|v\rangle=c$ for all unit vectors $|v\rangle$ within any eigenspace of $\rho_A$, and all cross-eigenspace matrix elements of $X$ must vanish. To determine the action of $X$ within a potentially degenerate eigenspace, we invoke the complex polarization identity. For any two orthonormal vectors $|x_1\rangle$ and $|x_2\rangle$ within the same eigenspace, the off-diagonal matrix element evaluates to
\[
\langle x_1|X|x_2\rangle = \frac{1}{4}\sum_{k=0}^{3} i^k \langle v_k|X|v_k\rangle\, ,
\]
where $|v_k\rangle = |x_1\rangle + i^k|x_2\rangle$. As the eigenspace is a linear subspace, each unnormalized superposition $|v_k\rangle$ resides entirely within it. Since $|x_1\rangle$ and $|x_2\rangle$ are orthonormal, each $|v_k\rangle$ has a squared norm of $2$. Moreover, since the expectation value of $X$ for any normalized vector in this space is $c$, it follows that $\langle v_k|X|v_k\rangle = 2c$ for all $k \in \{0, 1, 2, 3\}$. Substituting these values into the polarization identity then yields
\[
\langle x_1|X|x_2\rangle = \frac{1}{4}\left(2c + i(2c) - 2c - i(2c)\right) = 0\, .
\]
This demonstrates that $X$ acts as the scalar $c$ times the identity operator on each eigenspace, hence $X = c\mathds{1}_A$. Finally, since $X \in \text{Herm}_0(\H_A)$ is traceless, it follows that $c = 0$, thus $X = 0$, as desired.
\eprf

With Lemma~\ref{lem:span} established, the proof of our main result is straightforward. 

\bprf[Proof of Theorem~\ref{MXT71}]
Assume the NSIT condition in Eq.~\eqref{NSIT57} is satisfied for every projective measurement Alice and Bob can perform, and suppose $\rho_A \neq \mathds{1}_A/d_A$. It then suffices to prove that $\E$ is necessarily a discard-and-prepare channel. For this, we first restrict our attention to arbitrary binary projective measurements on $A$ of the form $\{P, \mathds{1}_A - P\}$, where $P$ is a rank-one projector. As NSIT is assumed for all possible projective measurements Alice and Bob can perform, Eq.~\eqref{NSIT57} yields
\[
\Tr \left( \E(P\rho_A P + (\mathds{1}_A - P)\rho_A(\mathds{1}_A - P)) Q \right) = \Tr \left( \E(\rho_A) Q \right)
\]
for all projectors $Q$ on $\H_B$. Since projectors span the space of Hermitian operators on $\H_B$, the non-degeneracy of the Hilbert-Schmidt inner product then implies
\[
\E \left( P\rho_A P + (\mathds{1}_A - P)\rho_A(\mathds{1}_A - P) \right) = \E(\rho_A)\, .
\]
By rearranging terms and utilizing the linearity of $\E$, we obtain $\E( \Delta_P(\rho_A)) = 0$ for all rank-one projectors $P$ on $\H_A$, where $\Delta_P(\rho_A)$ is the nonselective L\"{u}ders disturbance defined above.  Since by assumption $\rho_A \neq \mathds{1}_A/d_A$, Lemma~\ref{lem:span} guarantees that the real linear span of these disturbances $\Delta_P(\rho_A)$ constitutes the full space of traceless Hermitian operators on $\H_A$, namely $\text{Herm}_0(\H_A)$. As the linear map $\E$ vanishes on this spanning set of $\text{Herm}_0(\H_A)$, it follows that $\E(X) = 0$ for every $X \in \text{Herm}_0(\H_A)$. 

Now let $\omega_A$ be an arbitrary density operator on $\H_A$. We can uniquely decompose $\omega_A$ as
\be \label{TRXLESSDCMP67}
\omega_A = \frac{\mathds{1}_A}{d_A} + X_0\, ,
\ee
where $X_0 \in \text{Herm}_0(\H_A)$ is a traceless Hermitian operator. Applying the channel $\E$ to both sides  of Eq.~\eqref{TRXLESSDCMP67} yields
\[
\E(\omega_A) = \E\left(\frac{\mathds{1}_A}{d_A}\right) + \E(X_0) = \E\left(\frac{\mathds{1}_A}{d_A}\right)
\]
for all density operators $\omega_A$ on $\H_A$. Setting $\sigma_B = \E(\mathds{1}_A/d_A)$, it follows that $\E$ maps every input state to a fixed, constant output state $\sigma_B$, which is precisely the definition of a discard-and-prepare channel. This completes the proof.
\eprf

\begin{remark} \label{RMX17}
For a given state $\rho_A \neq \mathds{1}_A/d_A$ and an CPTP map $\E$ that is not a discard-and-prepare channel, a pair of binary measurements $\{P, \mathds{1}_A - P\}$ and $\{Q, \mathds{1}_B - Q\}$ violating NSIT can be explicitly constructed as follows:
\begin{enumerate}
    \item \emph{Construct Alice's measurement:} Calculate the nonselective L\"{u}ders disturbance $\Delta_P(\rho_A) = [P, [P, \rho_A]]$ for rank-one projectors $P =\dyad{v}{v}$. As the span of all such disturbances covers the entire space of traceless Hermitian operators (cf. Lemma~\ref{lem:span}), and $\E$ is not a discard-and-prepare channel, there is guaranteed to exist a unit vector $|v\rangle$ such that the output matrix $Y_P = \E(\Delta_P(\rho_A))$ for $P=\dyad{v}{v}$ is non-zero. Let Alice's measurement be the binary PVM $\{P, \mathds{1}_A - P\}$ for this specific $P$.
    \item \emph{Construct Bob's measurement:} The matrix $Y_P$ is Hermitian since $\E$ is necessarily Hermitian-preserving, and it is traceless since $\E$ is trace-preserving and $\Delta_P(\rho_A)$ is traceless. Since $Y_P$ is non-zero and traceless, it must possess at least one non-zero real eigenvalue $y_0 \neq 0$. Let $|y\rangle$ be the normalized eigenvector corresponding to $y_0$. We then set Bob's measurement to be $\{Q, \mathds{1}_B - Q\}$, where $Q = |y\rangle\langle y|$. 
\end{enumerate}
By evaluating the NSIT condition~\eqref{NSIT57} for these specific projectors, the difference between the measured and unmeasured marginals for Bob's outcome $Q$ reduces exactly to $\Tr[Y_P Q] = \langle y | Y_P | y \rangle = y_0$. Since $y_0 \neq 0$, NSIT is explicitly violated.
\end{remark}

\begin{remark} 
We note that the proof of Theorem~\ref{MXT71} only invokes the linearity of the map $\E$, without ever utilizing the fact that $\E$ is assumed to be completely positive and trace-preserving (CPTP). However, as the above procedure for determining measurements violating NSIT does in fact require $\E$ to be Hermitian-preserving and trace-preserving (HPTP), there exists the possibility of extending our results to dynamics described by HPTP maps. Such maps have been considered in effective descriptions of reduced open-system dynamics, including certain non-Markovian settings and dynamics involving initial system-environment correlations~\cite{Pechukas_1994,Breuer_2009,Rivas_2014}. These notions are closely related in such descriptions, but they are not equivalent. In particular, initial correlations do not by themselves imply a non-CP reduced evolution~\cite{Schmid_2019}. Moreover, a general HPTP map need not be positive on the full state space~\cite{Pechukas_1994}. Accordingly, the construction of NSIT-violating measurements as outlined above only retains such an interpretation when the initial state and the relevant post-measurement inputs lie in a compatibility domain on which an HPTP map $\E$ produces valid output states. Outside such a domain, the NSIT condition~\eqref{NSIT57} remains an algebraic statement rather than a comparison of physical probabilities.
\end{remark}

\section{Depolarizing dynamics}

As an illustration of our results, let us consider a sequential measurement scenario where a single qubit is initially prepared in the pure state $\rho_A = |0\rangle\langle0| = \frac{1}{2}(\mathds{1}_A + \sigma_z)$. We take the dynamics between measurements to be governed by a depolarizing channel given by
\[
\E(X) = (1-p)X + p\Tr(X)\frac{\mathds{1}_B}{2}\, ,
\]
where $p \in [0, 1]$ is the depolarizing parameter. 

Following Step 1 of Remark~\ref{RMX17}, we must choose a rank-one projector $P$ for Alice that does not commute with $\rho_A$. As such, let Alice measure in the $x$-basis by choosing $P = |+\rangle\langle+| = \frac{1}{2}(\mathds{1}_A + \sigma_x)$. The nonselective L\"{u}ders disturbance induced by this measurement is then
\[
\Delta_P(\rho_A) = \rho_A - P\rho_A P - (\mathds{1}_A-P)\rho_A(\mathds{1}_A-P) = \frac{1}{2}\sigma_z\, .
\]
Passing this disturbance through the channel yields the output matrix $Y_P$. Since $\Delta_P(\rho_A)$ is traceless, the trace term in the depolarizing channel vanishes, which yields
\[
Y_P = \E(\Delta_P(\rho_A)) = \E\left(\frac{1}{2}\sigma_z\right) = \frac{1-p}{2}\sigma_z\, .
\]

For Step 2, we must construct Bob's measurement $Q = |y\rangle\langle y|$ from the normalized eigenvector $|y\rangle$ corresponding to a non-zero eigenvalue $y_0 \neq 0$ of $Y_P$. The matrix $Y_P = \frac{1-p}{2}\sigma_z$ has eigenvalues $\pm \frac{1-p}{2}$. Assuming $p < 1$, we select the positive eigenvalue $y_0 = \frac{1-p}{2}$, which corresponds to the eigenvector $|0\rangle$. In such a case, the algorithm dictates Bob's measurement to be the binary projective measurement $\{Q, \mathds{1}_B-Q\}$ with $Q = |0\rangle\langle0|$. 

As guaranteed by Remark~\ref{RMX17}, the difference between Bob's measured and unmeasured marginals evaluates exactly to 
\[
\Tr(Y_P Q) = y_0 = \frac{1-p}{2}\, .
\]
It then follows that for all $p < 1$ the trace equality \eqref{NSIT57} fails and NSIT is explicitly violated. 

Furthermore, when $p = 1$ the channel $\E$ is completely depolarizing, mapping every input state to the maximally mixed state, making it a discard-and-prepare channel. In such a case we have $Y_P = 0$, hence there is no non-zero eigenvalue of $Y_P$ to select in Step 2 for the construction of Bob's violating measurement. Such a breakdown of Step 2 is in accordance with Theorem~\ref{MXT71}, which implies that when $\E$ is completely depolarizing, NSIT is universally satisfied, and no such measurements violating NSIT can exist.

\section{Dephasing dynamics}

To further illustrate the universality of Theorem~\ref{MXT71}, we consider a sequential measurement scenario where the open-system dynamics are governed by a general dephasing channel. Unlike the depolarizing channel, a dephasing channel preserves classical populations (diagonal elements) while suppressing quantum coherences (off-diagonal elements). As such, a dephasing channel is never a discard-and-prepare channel, thus Theorem~\ref{MXT71} guarantees that NSIT-violating measurements must exist for dephasing dynamics whenever the initial state is not maximally mixed. 

For a concrete realization, let the initial state of a single qubit be the coherent superposition $\rho_A = |+\rangle\langle+| = \frac{1}{2}(\mathds{1}_A + \sigma_x)$, and suppose the dynamics between measurements are governed by the dephasing channel in the $z$-basis, which is given by 
\[
\E(X) = \left(1-\frac{\gamma}{2}\right)X + \frac{\gamma}{2}\sigma_z X \sigma_z\, ,
\]
where $\gamma \in [0,1]$ is the dephasing parameter. Under this map, the Pauli operators evolve as $\E(\sigma_x) = (1-\gamma)\sigma_x$, $\E(\sigma_y) = (1-\gamma)\sigma_y$, and $\E(\sigma_z) = \sigma_z$.

Following Step 1 of Remark~\ref{RMX17}, in order to violate NSIT Alice must choose a measurement that disturbs the state $\rho_A$. If Alice were to measure exclusively in the $z$-basis, the nonselective L\"{u}ders disturbance would be purely off-diagonal, which would be completely erased in the limit of complete dephasing ($\gamma=1$). To ensure the disturbance survives, Alice can measure in a tilted basis by choosing the rank-one projector 
\[
P = \frac{1}{2}\left(\mathds{1}_A + \frac{1}{\sqrt{2}}\sigma_x + \frac{1}{\sqrt{2}}\sigma_z\right)\, .
\]
As the nonselective L\"{u}ders disturbance induced by this projector evaluates to 
\[
\Delta_P(\rho_A) = \rho_A - P\rho_A P - (\mathds{1}_A-P)\rho_A(\mathds{1}_A-P) = \frac{1}{4}\sigma_x - \frac{1}{4}\sigma_z\, ,
\]
we have that the operator $Y_P$ corresponding to Step 2 of Remark~\ref{RMX17} is given by
\[
Y_P = \E(\Delta_P(\rho_A)) = \frac{1-\gamma}{4}\sigma_x - \frac{1}{4}\sigma_z\, .
\]
Since $Y_P\neq 0$ for every value of the dephasing parameter $\gamma\in [0,1]$, it follows that there exists a violating measurement $\{Q,\mathds{1}_B-Q\}$ for Bob, where $Q = |y\rangle\langle y|$ is constructed from the normalized eigenvector $\ket{y}$ corresponding to a non-negative eigenvalue $y_0$ of the traceless operator $Y_P$. Selecting the positive eigenvalue yields
\[
y_0 = \frac{\sqrt{1 + (1-\gamma)^2}}{4}\, ,
\]
which is a monotonically decreasing function of the dephasing parameter $\gamma$. Crucially, since $y_0 \neq  0$ for all $\gamma \in [0,1]$, the NSIT condition~\eqref{NSIT57} is strictly violated for every possible strength of system-environment interaction. At $\gamma=0$ (unitary evolution), the violation evaluates to $\sqrt{2}/4$. In the extreme limit of complete dephasing ($\gamma=1$), the off-diagonal coherences are completely destroyed, yet the measurement footprint is preserved entirely within the classical populations, yielding a non-zero violation of $1/4$. This explicitly confirms that complete dephasing does not by itself erase the statistical footprint of an invasive measurement.

\section{Concluding Remarks}

Our main result provides a complete classification of universal NSIT in the two-time sequential measurement scenario considered here. A maximally mixed initial state is unchanged by Alice's nonselective projective measurement, while a discard-and-prepare channel makes the output state independent of the input. Outside these two cases, Theorem~\ref{MXT71} and Lemma~\ref{lem:span} show that binary projective measurements violating NSIT can always be constructed, as illustrated by the depolarizing and dephasing examples.

In the context of quantum causal inference~\cite{Costa_2016,Allen_2017}, recent studies have shown that there exist bipartite Pauli correlations between systems of qubits which are consistent with both spacelike separated measurements and sequential measurements~\cite{ZPTGVF18,song23,Song_2025,LiuX_2025}. Therefore, it is not possible to infer from the correlations obtained in such experiments whether or not the associated Pauli measurements were performed in sequence or in parallel. Our results suggest that intervention-based NSIT tests may provide additional information about this ambiguity by comparing Bob's statistics with and without Alice's earlier measurement. Determining the precise scope of this possible application is left for future work.

\emph{Acknowledgments.---} J.F. is supported by the National Natural Science Foundation of China (W2632013), and by the Hainan Provincial Natural Science Foundation of China (126MS0010). J.F. would also like to thank Naim Elias Comar for fruitful discussion. W.Y. is supported by National Natural Science Foundation of China (62475062), and the Humboldt Research Fellowship Programme for Experienced Researchers (CHN-1218456-HFST-E).

\emph{AI usage declaration.---} During the preparation of this manuscript, the authors utilized Google Gemini for language refinement, and as an interactive tool for the exploration of analytical methods and the preliminary structural organization of mathematical proofs. The authors rigorously reviewed and edited all generated text and assume full responsibility for the content and technical accuracy of this work.

\bibliography{references}

\end{document}